\documentclass{WileyMSP-template} 
\usepackage[numbers,sort&compress]{natbib} 
\usepackage{bm}
\usepackage{xcolor} \usepackage{soul} \usepackage[normalem]{ulem} \usepackage{amsmath,amsfonts,amssymb} \usepackage{physics} \usepackage{empheq} \usepackage{extarrows} \usepackage{multirow} \usepackage{mathrsfs} \usepackage{gensymb} \usepackage{verbatim} \usepackage[scr=boondoxo,scrscaled=1.05]{mathalfa} \DeclareMathAlphabet\mathrsfso{U}{rsfso}{m}{n} \def\e{\mathcal{E}}   \def\ee{\textrm{e-e}} \def\eph{\textrm{e-ph}}   \usepackage[colorlinks = true, linkcolor = magenta,anchorcolor = magenta, citecolor = magenta,filecolor = magenta, urlcolor = magenta]{hyperref} \usepackage{tikz} \usepackage{orcidlink} \usepackage{xr}  \definecolor{myred}{RGB}{210, 4, 45} \newcommand{\stkout}[1]{\ifmmode\text{\sout{\ensuremath{#1}}}\else\sout{#1}\fi} 

\newcommand{\XYZ}[2]{{\iffalse {#1} \fi}{#2}} 

\begin{document} 
\pagestyle{fancy} 
\rhead{\includegraphics[width=2.5cm]{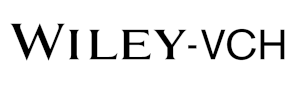}} \title{Origin, strength, and speed of the nonlinear optical response of transparent conducting oxides to single-cycle optical pulses} \maketitle 

\author{Ieng-Wai Un*\,\orcidlink{0000-0002-7156-0019}} \author{Subhajit Sarkar\,\orcidlink{0000-0002-7260-5100}} \author{Yonatan Sivan\,\orcidlink{0000-0003-4361-4179}} 



\begin{affiliations}
\justifying
\noindent Ieng-Wai Un\\
Key Laboratory of Atomic and Subatomic Structure and Quantum Control (Ministry of Education), Guangdong Basic Research Center of Excellence for Structure and Fundamental Interactions of Matter, School of Physics, South China Normal University, Guangzhou 510006, China.\\
Guangdong Provincial Key Laboratory of Quantum Engineering and Quantum Materials, Guangdong-Hong Kong Joint Laboratory of Quantum Matter, South China Normal University, Guangzhou 510006, China.\\
School of Electrical and Computer Engineering, Ben-Gurion University of the Negev, Beer Sheva, 8410501, Israel.\\
Email Address: iengwai@m.scnu.edu.cn

\noindent Subhajit Sarkar\\
Department of Physics, School of Natural Sciences, Shiv Nadar Institution of Eminence Deemed to be University, Delhi-NCR, NH91, Tehsil Dadri, Greater Noida, Uttar Pradesh 201314, India.\\
School of Electrical and Computer Engineering, Ben-Gurion University of the Negev, Beer Sheva, 8410501, Israel.

\noindent Yonatan Sivan\\
School of Electrical and Computer Engineering, Ben-Gurion University of the Negev, Beer Sheva, 8410501, Israel.

\end{affiliations}


\keywords{single-cycle pulses, transparent conducting oxides, optical nonlinearity, quantum coherence, thermalization}

\begin{abstract}
\justifying
\noindent We present a first-principles study of the nonlinear optical response of transparent conducting oxides at the nanoscale due to excitation by intense, extremely short pulses based on a density matrix framework. We identify a strong ($O(1)$) thermal nonlinearity, which is complemented with stimulated emission and excited-state absorption; it yields a cumulative permittivity change decorated by quantum coherent oscillations. Further, rigorous calculations under far-from-equilibrium conditions show that electron-electron thermalization occurs within a few femtoseconds, supporting interpretations of high-harmonic generation measurements and in agreement with a generalization of Fermi liquid theory.

\end{abstract}


\justifying
\section{Introduction}
The interaction of intense single- or sub-cycle optical pulses with solids is of importance to our understanding of many-body interactions, coherence, and dynamics in (electron) systems far from equilibrium\XYZ{; it}{. It} also has practical importance, e.g., as a potential source for single-electron emission from metal tips~\cite{Kruger_review}, and as a source for high-harmonic generation. 

So far, most studies involved coherent extremely short pulse interactions with dielectrics~\cite{Stockman-current-in-dielectrics,Stockman-current-in-dielectrics2}, (intrinsic) semiconductors~\cite{Huber_QDots_2009,Nirit_2022,Koch_Kira_Huber_2022}, and less so with noble metals~\cite{Stockman-metal-film-fs,Yabana_ACS_Photonics_2019,Marinica-quantum-dimer}. The accompanying theory usually relied on Time-Dependent Density Functional Theory (TD-DFT)~\cite{Gross_TDDFT,TDDFT_book,Marinica-quantum-dimer,Yabana_PRA_T2_diamond} or the Maxwell-Bloch equations~\cite{Meier_Koch_PRB_2008,Peschel_Maxwell_Bloch} (or equivalently, the Density Matrix (DM) formulation~\cite{Krieger_1987,Boyd-book,Vampa_PRL_2014})\XYZ{; more}{. More} recently, advanced (and \XYZ{}{computationally much} heavier) approaches such as finite-temperature DFT~\cite{colombier_2022} and many-body perturbation theory~\cite{colombier_2024_2,colombier_2024} were applied to extremely short pulse interactions with dielectrics or gases~\cite{Brezinova_2015}. In these works, most attention was given to the early stages of the dynamics, to the coherence between electrons and photons and the associated nonlinearities, as well as to quantum and transport phenomena in nanometric structures. This involved a meticulous calculation of the various electron wavefunctions, self-energies, and then the transition matrix elements in real space. Overall, the predictions from these approaches are in good agreement with measurements. 

In contrast, the incoherent nonlinear optical response, i.e., intensity-dependent permittivity changes, received little attention; some exceptions are Refs.~\cite{Schultze_Kerr_SiO2,colombier_2024_2,Herbst_fieldoscopy_AdvSci_2026}. Moreover, effects such as excited-state absorption and stimulated emission were rarely studied\XYZ{ because of the difficulty in monitoring}{, largely because} the non-equilibrium electron population dynamics \XYZ{}{are difficult to measure experimentally or to monitor} using the \XYZ{}{theoretical} techniques mentioned above\XYZ{; instead}{. Instead}, population dynamics is usually treated with incoherent rate equations. Moreover, the later stages of the dynamics, namely, the thermalization via electron-electron ({\em e}-{\em e}) interactions and relaxation via electron-phonon ({\em e}-{\em ph}) interactions are usually treated by the simplified Relaxation Time Approximation~\cite{Ashcroft-Mermin,Krieger_1987,Ullrich_PRL_2006,Yabana_PRB_T2} or assuming thermal populations~\cite{colombier_2024,colombier_2024_2}; a rare exception~\cite{Turkowski_PRB_2022} studied the effect of {\em e}-{\em e} interactions on the harmonic generation, but not the effect of {\em e}-{\em ph} and electron-impurity ({\em e}-{\em imp}) interactions, nor the effect of the latter on the permittivity dynamics; similarly, electron collision dynamics were studied in insulators under the conditions of thermal equilibrium~\cite{colombier_2024,colombier_2024_2}. As a result, there is not yet a detailed understanding of thermalization and relaxation dynamics in the far-from-equilibrium conditions created by the extremely short pulses. A prime example is in the field of high harmonic generation~\cite{Yabana_PRB_T2,Moloney_Koch_PRL_HHG_propagation_2020,Yabana_PRA_T2_diamond} where there are disagreements about the proper value that has to be assigned to the dephasing rates (typically termed as $T_2$ in the context of atomic physics). Specifically, while originally values of (at least) a few 10's of femtosecond were used, some examples showed that good match with experimental data requires values as short as $1$ fs~\cite{Vampa_PRL_2014,Huber_T2_1fs,Goulielmakis_T2_1fs,Luu_PRB_2016_T2_1fs,Ghimire_T2_1fs,Lu_PRA_2016_T2_1fs}. On the other hand, more recent works demonstrated that experimental data can be matched with longer $T_2$ values (i.e., 10's-100's femtoseconds) if full $k$-space modelling~\cite{Yabana_PRA_T2_diamond} and propagation effects are taken into account, e.g., in diamond~\cite{Yabana_PRA_T2_diamond} or in GaAs~\cite{Moloney_Koch_PRL_HHG_propagation_2020}.

To address these open questions, we combine the DM formulation with a rigorous treatment of the (many-body) thermalization and relaxation~\cite{delFatti_nonequilib_2000,Italians_hot_es,GdA_hot_es,Sarkar-Un-Sivan-Dubi-NESS-SC,Un-Sarkar-Sivan-LEDD-I,Un-Sarkar-Sivan-LEDD-II}. We employ the formulation to study the nonlinear optical response of transparent conducting oxides (TCOs; aka low electron density Drude materials~\cite{Un-Sarkar-Sivan-LEDD-I,Un-Sarkar-Sivan-LEDD-II} or $p$-doped metals~\cite{Zayats_CuS}). TCOs have recently attracted the attention of the nonlinear optics community due to their extremely high nonlinearities. Indeed, permittivity changes of several 100's of percent were observed from thin films with pulses a few 10s-100s femtosecond long~\cite{Boyd_NLO_ENZ_ITO,Shalaev_Faccio_NLO_ENZ,Yang-HHG-CdO-natphys-2019,Exeter_Nat_Comm_2021,Ellenbogen-Minerbi-ITO,de_Leon_Mexicans_2022,Shalaev_Faccio_NLO_ENZ,Sapienza_2022,H_Lee_Nanophotoics_2023}. This has made TCOs promising candidates for photonic time crystals~\cite{Pendry_time_variations,Khurgin_photonic_time_crystal_vs_4Wmix}. Originally, the strong nonlinearity was associated with the zero permittivity crossing point~\cite{Boyd_NLO_ENZ_ITO,Boyd_Nat_Phot_2018,Reshef-Boyd_review-2019}, and then, with the conduction band non-parabolicity~\cite{Secondo_OME_2019}; more recently, we showed that a significant contribution also comes from the increased collision rates induced by the heating~\cite{Un-Harcavi-Sivan-LEDD-Viewpoint} and that the heating can be far more significant than expected originally~\cite{Un-Sarkar-Sivan-LEDD-I}, even exceeding the Fermi temperature ($\sim 10,000$ K)~\cite{Blemker_ITO_NPs_hot}.

However, to date, there is no comprehensive understanding of the interactions of TCOs with extremely short pulses. As a follow-up of earlier work~\cite{Un-Sarkar-Sivan-LEDD-II,Scalora-ITO-2025}, this work aims to close this knowledge gap by providing a {\em non-phenomenological} study of this problem and to match the rigorous model results to those from simpler ones. To do that, we first employ the DM formulation to study the dynamics of electron population and coherence, as well as of the macroscopic polarization and local field. We elucidate the relative importance of the various electron interactions for the nonlinear response\XYZ{ and}{. We also} find that the response to extremely short pulses has many similarities to the response to longer pulses, e.g., the overall strength, the cumulative nature of the response, and the unique large changes to the real part of the permittivity. Nevertheless, the response to extremely short pulses has a more complex electric field dependence due to stimulated emission and excited state absorption, and due to the effect of coherence between the various electron states on the permittivity dynamics. Nevertheless, we also show that, surprisingly, one can reproduce the permittivity dynamics quite accurately with a far simpler thermal model, which simplifies the modelling and understanding of this problem drastically.

Moreover, our treatment provides a rigorous insight into the accelerated collision rates and the associated nonlinearity decay in the context of extremely short pulses that goes beyond the most advanced treatments for noble metals~\cite{delFatti_nonequilib_2000} and insulators~\cite{colombier_2024,colombier_2024_2}. Specifically, we find that the thermalization/collision time can be shortened to $\sim 10$ femtoseconds for TW/cm$^2$ intensities, and that its scaling with $|E|^2$ is linear (for the range of intensities studied here). This provides a first-of-its-kind calculation that is inline with the debated few femtosecond thermalization times used previously to explain high harmonic generation measurements~\cite{Yabana_PRB_T2,Moloney_Koch_PRL_HHG_propagation_2020,Yabana_PRA_T2_diamond,Vampa_PRL_2014,Huber_T2_1fs,Goulielmakis_T2_1fs,Luu_PRB_2016_T2_1fs,Ghimire_T2_1fs,Lu_PRA_2016_T2_1fs}. We also provide a {\em unique} analytic generalization of Fermi's Liquid Theory (FLT) for the {\em e}-{\em e} collision (thermalization) rate for high temperatures and far-from-equilibrium conditions, which reproduces the rigorous numeric calculation very well. This is far from being a trivial result, since FLT is derived assuming the electron distribution is thermal, whereas the distribution becomes strongly non-thermal under the investigated conditions. This result would be useful in studying thermalization dynamics in high-harmonic generation and attosecond science.

Finally, we compare our theoretical predictions to the measurements of the nonlinear optical response of TCOs to extremely short pulses~\cite{Shalaev_Segev_nanophotonics_2023}, as well as to alternative (yet more phenomenological) modelling.

\section{Formulation}
We solve the DM equations~\cite{Boyd-book,Meier_Koch_PRB_2008,Peschel_Maxwell_Bloch} for a popular TCO, namely, Indium Tin Oxide (ITO) in momentum space (see details in Reference~\cite{single_cycle_nlty_Article}). Electron transition moments are described within the dipole approximation assuming an infinitely deep potential is formed by the particle geometry~\cite{Ropers_PRB_2011,Govorov_1,Khurgin-Levy-ACS-Photonics-2020}; we account for the conduction band non-parabolicity (Kane formula~\cite{Kane-quasilinear}) by writing the transition dipole moment in terms of the momentum operator using the $p$-$r$ relationship~\cite{p-r-relation} and replacing the electron energy $\mathcal{E}$ by $\mathcal{E}(1 + C \mathcal{E})$, namely, $- e \bra{\psi_{{\bf k}}}\hat{{\bf r}}\ket{\psi_{{\bf k}'}} = ie\hbar\bra{\psi_{{\bf k}}}\hat{{\bf p}}\ket{\psi_{{\bf k}'}}/( \e_{{\bf k}}(1 + C \e_{{\bf k}}) - \e_{{\bf k}'}(1 + C \e_{{\bf k}'}))$, where $C$ is the non-parabolicity parameter see Reference~\cite{single_cycle_nlty_Article}, Section~II B. The collision rates, $\eta_{\ee}$ and $\eta_{\eph}$ are treated {\em rigorously in ${\bf k}$-space} under the Thomas-Fermi and deformation potential approximations, respectively, as in Refs.~\cite{delFatti_nonequilib_2000,Italians_hot_es,GdA_hot_es,Sarkar-Un-Sivan-Dubi-NESS-SC,Un-Sarkar-Sivan-LEDD-I,Un-Sarkar-Sivan-LEDD-II}.

Since the pulse duration is shorter than the material response time (i.e., the collision rates), we a-priori refrain from defining a permittivity, and instead, compute the polarization density via tracing the dipole matrix element and the DM off-diagonal terms~\cite{Boyd-book}. 

\section{Configuration}
\begin{figure}[h]
\centering
\includegraphics[width=10cm]{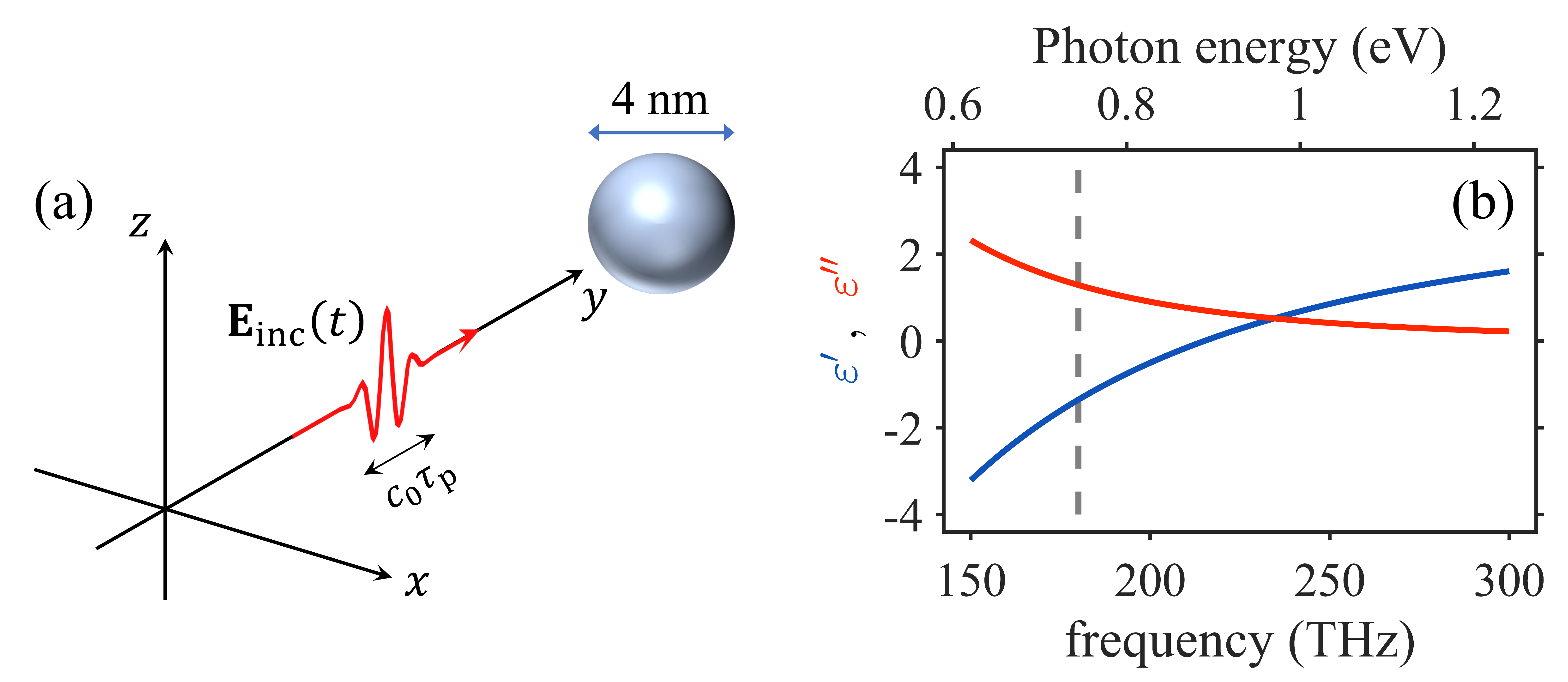}
\caption{(a) Schematic diagram of the setup: A pump pulse is incident on an ITO NP with a diameter of 4 nm embedded in air. (b) The real (blue solid line) and imaginary (red solid line) parts of the ITO permittivity at room temperature. The gray dashed line indicates the resonance frequency ($\sim 180$ THz).}
\label{fig:scheme_eps}
\end{figure}

In order to focus on the features unique to the dynamics of conduction electrons, we chose a configuration of a few nm ITO sphere (Figure~\ref{fig:scheme_eps}(a)), using the material parameters extracted from earlier experiments, as described in Refs.~\cite{Un-Sarkar-Sivan-LEDD-I,Un-Sarkar-Sivan-LEDD-II}. This approach has multiple advantages. First, as seen in Figure~\ref{fig:scheme_eps}(b), the dipolar resonance associated with the dipole mode of the nanosphere appears at $\hbar \omega \approx 0.7$eV, i.e., at wavelengths far from the threshold for interband transitions (thus, making our ``intraband-only'' treatment sufficient), as well as longer than the epsilon-near-zero point (thus, disentangling the dynamics from the associated pure electrodynamic effects associated with such points). Second, the electric field is uniform for such a system, thus entailing significant computational simplification and ensuring that dephasing due to spatial variations of the electromagnetic fields~\cite{Yabana_PRA_T2_diamond,Moloney_Koch_PRL_HHG_propagation_2020} does not play a role. 

The incident pulse is assumed to have a Gaussian envelope set to ${\bf E}_\textrm{inc}(t) = \hat{{\bf z}} E_0 e^{-2\ln 2 (t/\tau_p)^2}$ $\cos(\omega_0 t)$, where $E_0$ is its peak amplitude, $\tau_p = 5$ femtoseconds is its duration, and $\omega_0/2\pi = 175$ THz ($\sim 0.72$ eV or equivalently, a $\sim 6$ femtoseconds long cycle) is the central frequency. Since the size of the ITO nanoparticle (NP) is smaller than the pump wavelength, we employ the quasi-static approximation to calculate the field dynamics, i.e., we assume that the electric field inside the NP is uniform, i.e., ${\bf E}_\textrm{NP}(t) = \hat{{\bf z}}E_\textrm{NP}(t)$. In particular, the time-dependence of the electric field inside the NP is self-consistently related to the electric field of the incident pulse by (see details in Reference~\cite{single_cycle_nlty_Article})
\begin{align}\label{eq:Epump_to_Ein}
{\bf E}_\textrm{NP}(t) = \hat{{\bf z}} E_\textrm{NP}(t) = \dfrac{\hat{{\bf z}}}{\varepsilon_\infty + 2 \varepsilon_h} \left[3 \varepsilon_h E_\textrm{inc}(t) -\dfrac{P_\textrm{NP}(t)}{\varepsilon_0} \right].
\end{align}
Here, the first and second terms on the right-hand side of Equation~\eqref{eq:Epump_to_Ein} represent the non-dispersive and dispersive parts of the optical response of the NP, respectively.

\section{Results}
\subsection{Nonlinearity turn on}
\begin{figure}[h]
\centering
\includegraphics[width=1\columnwidth]{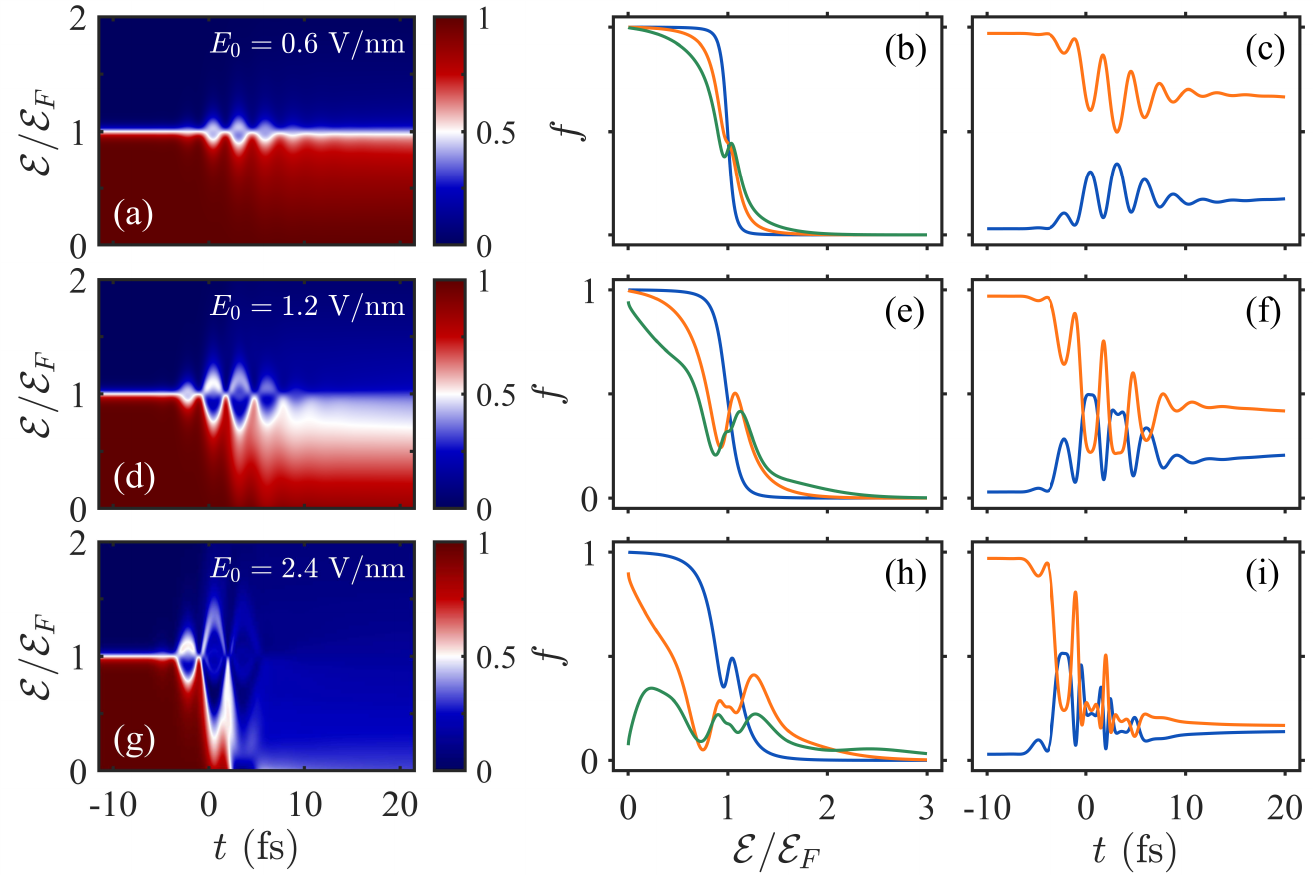}
\caption{(a) An energy-time map of the electron distribution dynamics, (b) fixed time snapshots of the electron distribution at $t = -2.5$ femtoseconds (blue), $0$ femtoseconds (orange), and $5$ femtoseconds (green), and (c) the electron distribution dynamics at $\e = \e_F \pm \hbar \omega_0/8$ (blue and orange) for $E_0 = 0.6$ V/nm. (d)-(f) and (g)-(i) show similar data as (a)-(c) for $E_0 =1.2$ V/nm and $2.4$ V/nm, respectively.}
\label{fig:fe_0.6_1.2_2.4Vm-1}
\end{figure}

We initially look at the electron dynamics due to illumination. Specifically, Figure~\ref{fig:fe_0.6_1.2_2.4Vm-1} (and Figure~S2
) show, respectively, the dynamics of the electron distributions and the excited electron density (Equation~(S27)
) for $E_0 = 0.6$ V/nm, $1.2$ V/nm (comparable to the field levels used in References~\cite{Shalaev_Segev_nanophotonics_2023,Herbst_fieldoscopy_AdvSci_2026}), and $2.4$ V/nm. While a strong deviation from thermal equilibrium is clearly visible in the electron distribution, fixed-time snapshots (Figures~\ref{fig:fe_0.6_1.2_2.4Vm-1}(b), (e), (h) and Figures~S2(b), (e), (h)
) show that the deviation does not resemble the non-thermal shoulders above the Fermi energy which are characteristic of metals under continuous wave and long pulse illumination (see e.g., Refs.~\cite{non_eq_model_Rethfeld,Italians_hot_es,Stoll_review,Dubi-Sivan,Dubi-Sivan-Faraday,Kumagai-ACS-phot-2023}); this is a signature of momentum conservation in the photon-e interactions~\cite{Govorov_1}. Moreover, those snapshots also reveal transient strong and oscillatory population inversion near the Fermi energy for sufficiently strong pump pulses (Figures~\ref{fig:fe_0.6_1.2_2.4Vm-1}(d)-(i)). The oscillation (seen more clearly in fixed energy cross sections, Figures~\ref{fig:fe_0.6_1.2_2.4Vm-1}(c), (f), (i) and Figures~S2(c), (f), (i)
) originates from the quantum coherence associated with the off-diagonal elements of the DM, which also manifests as oscillations in the e-photon interaction term~\cite{single_cycle_nlty_Article}. Notably, after the passage of the pump pulse, the population changes persist for the duration of the simulation ($\sim 20$ femtoseconds; see Figure~\ref{fig:fe_0.6_1.2_2.4Vm-1}(c), (f), and (i)) with no apparent rapid decay.

\begin{figure}[h]
\centering
\includegraphics[width=0.8\columnwidth]{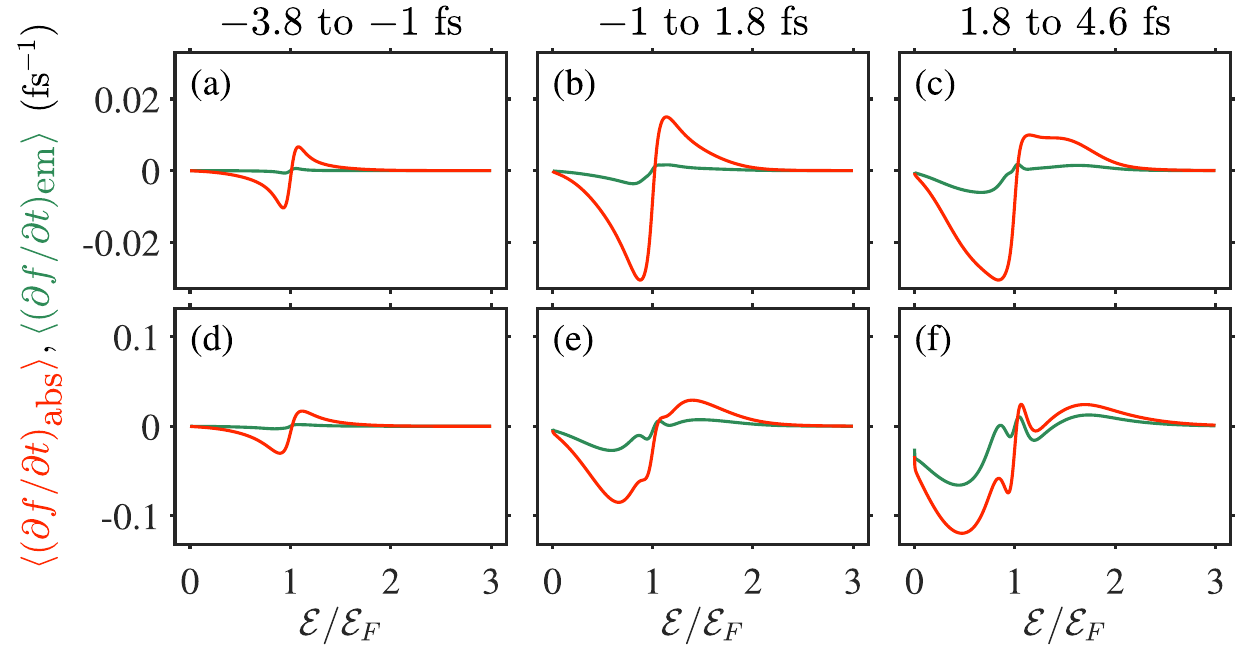}
\caption{(a)-(c) The local time average of the (stimulated) absorption (red) and (negative of the) stimulated emission (green) parts of the photon-e terms for $E_0 = 0.6 $ V/nm. The starting and ending points of the oscillation cycles are shown as titles. (d)-(f) The same as (a)-(c) for $E_0 = 1.2 $ V/nm.
}
\label{fig:dfdt_abs_emi_avg_0.6_1.2Vm-1_REF}
\end{figure}

This finding implies that the initial stage of the electron dynamics is dominated by photon-electron interactions, hence, it is fully coherent; this contrasts the case of longer pulses where electron collisions may occur simultaneously to photon-e interactions. Thus, we now perform a more detailed exploration of the photon-electron interactions by distinguishing (stimulated) absorption from stimulated emission. To circumvent the scrambling effect of the oscillatory nature of the associated terms, we look at their time average, namely,
$$
\left\langle\left(\frac{\partial f(\mathcal{E},t)}{\partial t}\right)_{\textrm{abs,em}}\right\rangle_{(t_i,\Delta t)} = \dfrac{1}{\Delta t} \int_{t_i}^{t_i  +\Delta t} \left(\dfrac{\partial f(\mathcal{E}, t)}{\partial t}\right)_{\textrm{abs,em}} \, dt',
$$ 
taken over every half cycle ($2\pi / 2 \omega_0 \sim 3$ femtoseconds), see Section~S1. 
Figure~\ref{fig:dfdt_abs_emi_avg_0.6_1.2Vm-1_REF} shows that while stimulated emission is much smaller than absorption for the early stages of the dynamics, it becomes comparable to absorption at the highest intensities studied; yet, it does not exceed it.  

\begin{figure}[h]
\centering
\includegraphics[width=1\columnwidth]{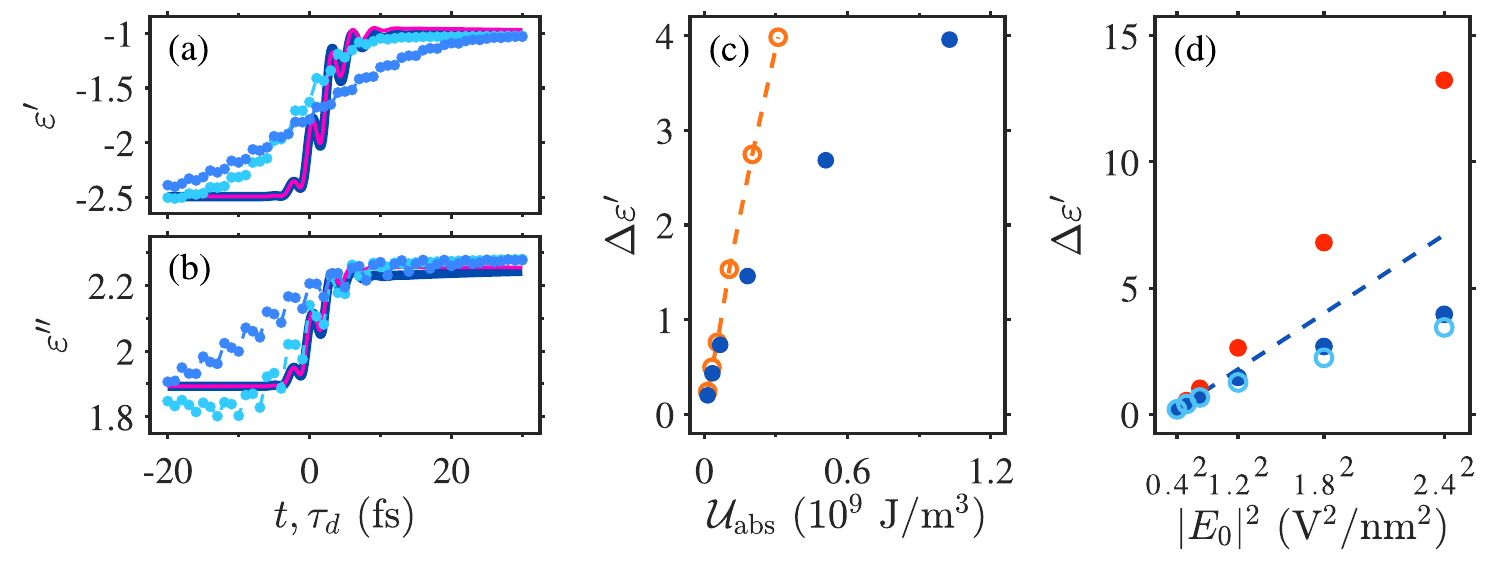}
\caption{Dynamics of the (a) real and (b) imaginary parts of the permittivity for a probe pulse with $\omega_\textrm{pr} = 2 \pi \times 160$ THz and $20$ fs (blue dots) and $5$ fs (cyan dots) durations. The blue and magenta lines represent the ANTH$\mathrsfso{E}$M permittivity (Equation~\eqref{eq:eps_ANTHEM_k})~\cite{Un-Sarkar-Sivan-LEDD-II} and the corresponding results from a purely thermal model~\eqref{eq:Te_coherent}, respectively. The peak field strength of the pump pulse is $E_0 = 1.2$ V/nm. (c) Maximal permittivity changes plotted vs. the absorbed energy density $\mathcal{U}_\mathrm{abs}$ with (blue filled circles) and without (orange unfilled circles) accounting for stimulated emission. (d) Maximal permittivity changes plotted vs. $|E_0|^2$ (blue filled circles). Blue open circles represent the contribution from spectral shift and broadening only, while red filled circles include the additional contribution from excited-state absorption. \XYZ{The red dashed lines in (c) and (d)}{The orange dashed line in (c) and the blue dashed line in (d)} represent the linear fit to $\Delta\varepsilon'$ for small fields.} \label{fig:permittivitty_dynamics}
\end{figure}

Now, to investigate the {\em macroscopic} optical response of the material arising from the microscopic dynamics, we perform a calculation that mimics the standard experimental procedure of a pump-probe measurement. Specifically, once we have determined the DM element dynamics for the single-cycle pump pulse, we compute the polarization induced by a weaker short probe (at variable central probe frequencies $\omega_\textrm{pr}$, duration, and pump-probe delay $\tau_d$) in the time domain. We then Fourier transform the induced polarization and divide it by the transform of the local electric field. The resulting coherent (non-adiabatic and) non-thermal response can be viewed as a type of ``instantaneous permittivity'', $\varepsilon(\omega_\textrm{pr},\tau_d)$. The results, shown in Figure~\ref{fig:permittivitty_dynamics}(a)-(b), reveal a rapid growth of the real and imaginary parts of $\varepsilon(\omega_\textrm{pr},\tau_d)$, reaching even $\sim 250\%$ relative changes; such strong nonlinearity is comparable to that observed in experiments and simulations for longer pulses of comparable fluence (e.g., that employed in the examples of Reference~\cite{Un-Sarkar-Sivan-LEDD-II}, $\sim 160$ GW/cm$^2$ $\cdot\ 30$ femtoseconds $\sim 5$ mJ/cm$^2$). Importantly, the response is not instantaneous, but rather it accumulates in time, and like the population, persists thereafter, indicating its absorptive nature. In that sense, the Drude nonlinearity we investigate here should not be approximated by a Kerr nonlinearity\XYZ{}{, which scales with the instantaneous intensity $|E(t)|^2$ and follows the temporal envelope of the optical pulse, see e.g., Reference~\cite{Schultze_Kerr_SiO2}}. \XYZ{Nevertheless, a Kerr nonlinearity can capture well the short rise time for the short pump pulse used.}{Nevertheless, for ultrashort excitation pulses, a Kerr-type description can provide a reasonable approximation of the initial rise of the nonlinear response, while it cannot describe the subsequent persistence of the induced permittivity change.}\footnote{\XYZ{}{Moreover, the Kerr effect represents an intrinsic optical nonlinearity determined by the material properties, whereas the thermo-optic response is a cumulative effect that depends not only on the material parameters (e.g., thermal diffusivity) but also on the device geometry and the optical and thermal properties of the surrounding environment. Therefore, an effective Kerr coefficient extracted from a thermo-optic response is generally configuration-dependent rather than a universal material parameter. This distinction has been discussed in detail in~\cite{Gurwich-Sivan-CW-nlty-metal_NP, Sivan-Chu-high-T-nl-plasmonics, IWU-Sivan-CW-nlty-metal_NP}.}} A closer look shows that the rise time of the extracted imaginary part of the ITO permittivity decreases with the probe pulse duration (see Figures~\ref{fig:permittivitty_dynamics}(a) and (b)), while the dynamics of the change of the real part is less sensitive to the probe duration. 

Remarkably, unlike the incoherent dynamics employed in Refs.~\cite{Un-Sarkar-Sivan-LEDD-I,Un-Sarkar-Sivan-LEDD-II}, the permittivity growth is decorated by the $1 / 2 \omega_0 \sim 3$ femtoseconds oscillation seen in the electron dynamics; thus, it is the clear signature of the coherence between the different electron states (i.e., the off-diagonal elements of the DM), as seen previously in TD-DFT simulations~\cite{Schultze_Kerr_SiO2}. 

Further insight is obtained by computing the ``instantaneous permittivity'' within the adiabatic non-thermal permittivity ($\mathrsfso{E}$) model (ANTH$\mathrsfso{E}$M), see Equation~(S12) 
and \XYZ{}{Equation~(5) in} Reference~\cite{Un-Sarkar-Sivan-LEDD-II}. \XYZ{This response can be viewed as}{The ``instantaneous permittivity'' is a time-dependent quantity obtained by evaluating the permittivity for the instantaneous electron distribution. In this sense, it corresponds to} the zero probe duration limit of the pump-probe curves; since ANTH$\mathrsfso{E}$M is computed here from the coherent dynamics of the electron population obtained from the DM equations, it is oscillatory as well. Finally, we also present the permittivity dynamics computed using a thermal model in which the evolution of an effective electron temperature $\tilde{T}_e$ (defined in Equation~(S20)
) is determined by balancing the heat exchange among photons, electrons, and the lattice, i.e.,
\begin{align}\label{eq:Te_coherent}
C_e(\tilde{T}_e) \dfrac{d \tilde{T}_e}{dt} = - G_\textrm{e-ph}(\tilde{T}_e - T_{ph}) +  \dot{{\bf P}}_\textrm{NP}(t) \cdot {\bf E}_\textrm{NP}(t).
\end{align}
Here, $C_e(\tilde{T}_e)$ is the temperature-dependent heat capacity of the electron subsystem~\cite[Equation~(25)]{Un-Sarkar-Sivan-LEDD-I}, $T_{ph}$ is the lattice temperature, $G_\textrm{e-ph}$ is the {\em e}-{\em ph} coupling coefficient (Equation~(26) in Reference~\cite{Un-Sarkar-Sivan-LEDD-I}), ${\bf P}_\textrm{NP}$ is the polarization density in the NP given by Equation~(S10)
, and $ \dot{{\bf P}}_\textrm{NP}(t) \cdot {\bf E}_\textrm{NP}(t)$ is the {\em net} absorption power density (i.e., it accounts for both absorption and stimulated emission). The electron temperature is concurrently used to determine the (thermal / Fermi-Dirac) electron distribution, and to compute the DM off-diagonal terms via Equation~(10b) of Reference~\cite{single_cycle_nlty_Article} as well as the associated relaxation rates, the polarization density $P_\textrm{NP}(t)$, and the corresponding local field via Equation~\eqref{eq:Epump_to_Ein}\footnote{Specifically, we solve Equation~(13) in Reference~\cite{single_cycle_nlty_Article} for the DM off-diagonal terms using the time evolution of the thermal distribution $f^T(\mathcal{E},\mu(\tilde{T}_e(t)),\tilde{T}_e(t))$. The results are then used to compute $P_\textrm{NP}$ via Equation~(14) of Reference~\cite{single_cycle_nlty_Article}. 
}. Remarkably, even without accounting for the non-thermal distribution~\cite{Dubi-Sivan,Dubi-Sivan-Faraday,Dubi-Sivan-MJs} and energy~\cite{de_Leon_Mexicans_2022,Un-Sarkar-Sivan-LEDD-II}, the permittivity obtained from the thermal model~\eqref{eq:Te_coherent} is nearly identical to the ANTH$\mathrsfso{E}$M result, despite the strong deviation from thermal equilibrium (Figure~\ref{fig:fe_0.6_1.2_2.4Vm-1}). We associate this success with the thermal nature of the absorption term, see Reference~\cite{Govorov_1}, Figure~\ref{fig:dfdt_abs_emi_avg_0.6_1.2Vm-1_REF} and Figure~2(a)-(c) of Reference~\cite{single_cycle_nlty_Article}, as well as to the extremely fast {\em e}-{\em e} interactions, see below. The thermal model also shows the $1 / 2 \omega_0$ oscillation, which now stems from the phase delay between the electric field and the current it induces ($\sim \dot{P}_\textrm{NP}$). This oscillation is usually averaged out for long pulses, see, e.g., Refs.~\cite{Jackson-book,Un-Sarkar-Sivan-LEDD-II}, in which case the absorption density reduces to $\omega_0 \varepsilon''(\omega_0) |{\bf E}_\textrm{NP}(t;\omega_0)|^2/2$. 

Now, in order to understand the strength of the nonlinear response (i.e., the permittivity changes), in Figure~\ref{fig:permittivitty_dynamics}(c) we plot the maximal value of the permittivity change\footnote{Since both real and imaginary parts grow upon illumination, we focus only on the real part.}, $\Delta \varepsilon(\tau_d) \equiv \varepsilon(\omega_\textrm{pr},\tau_d) - \varepsilon(\omega_\textrm{pr},-\infty)$, as a function of the time-integrated absorbed energy density, i.e., $\mathcal{U}_\mathrm{abs} \equiv \int^\infty_{-\infty} [\dot{{\bf P}}_\textrm{NP,abs}(t) \cdot {\bf E}_\textrm{NP}(t)] dt$, where $\dot{{\bf P}}_\textrm{NP,abs}$ is the time derivative of the absorption component of the polarization density (see Equation~(S14)
).

The result shows a clear sublinear scaling of $\Delta\varepsilon'$ with the actual absorption. For comparison, we also consider the case in which the density-matrix equations are solved without including stimulated emission\footnote{This is done by neglecting the equation of motion of the density matrix off-diagonal elements associated with stimulated emission (Equation~(S16b)
).}. In this case, the permittivity change scales linearly with $\mathcal{U}_\mathrm{abs}$. This indicates that the permittivity change has its roots in the change of the total energy of the electron system~\cite{Un-Sarkar-Sivan-LEDD-II}, while stimulated emission reduces the net nonlinear response.

This aspect of the nonlinear response can be further elucidated by decomposing the change in permittivity into distinct physical contributions, including modifications due to stimulated emission and absorption rates, as well as spectral shift and resonance broadening. In particular, the absorbed energy density associated with the contribution arising solely from spectral shift and broadening is obtained from integrating $\omega \varepsilon_\mathrm{lin}''(\omega)|\tilde{E}_\mathrm{NP}(\omega)|^2/2$ over the angular frequency, where $\varepsilon_{\textrm{lin}}''(\omega)$ is the imaginary part of the permittivity in the weak-field (linear) limit (Equation~(S11)
), and $\tilde{E}_\mathrm{NP}(\omega)$ is the Fourier transform of the local field. As shown in Figure~\ref{fig:permittivitty_dynamics}(d) and Figure~S3
, the increased imaginary part (due to the so-called quantum size effect) makes broadening dominate over the spectral shift, and leads to a sublinear scaling of the permittivity change with $|E_0|^2$ (in similarity to the case of longer pulses~\cite{Un-Sarkar-Sivan-LEDD-II}). Including the excited state absorption significantly enhances the permittivity change, resulting in a superlinear scaling with $|E_0|^2$. However, this enhancement is largely compensated by the stimulated emission, whose contribution scales sublinearly with $|E_0|^4$ (see Figure~S3
). Consequently, the overall permittivity change exhibits a sublinear dependence on $|E_0|^2$ (hence, a sub-cubic nonlinearity), only slightly exceeding that obtained from spectral effects alone. In that sense, the response of the few nm ITO particles differs from that of thicker ITO layers (for which resonance shifts are responsible for the sublinear growth of the temperature and the corresponding sub-cubic nonlinear response), and shows a similar response to that of Au NPs under CW illumination~\cite{Gurwich-Sivan-CW-nlty-metal_NP,IWU-Sivan-CW-nlty-metal_NP}. 

\subsection{Nonlinearity turn off (relaxation/(de)coherence time)}
\begin{figure}[h]
\centering
\includegraphics[width=1\columnwidth]{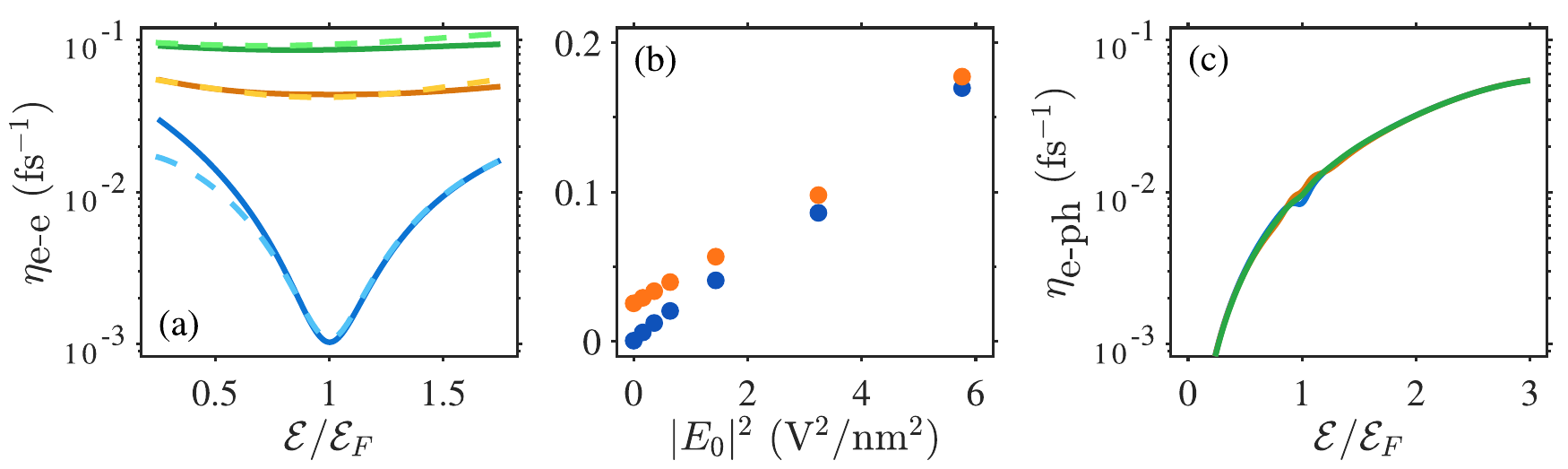}
\caption{(a) The {\em e}-{\em e} collision rate for $E_0 = 1.8$ V/nm at $t = -5$ femtoseconds (blue), $0$ femtoseconds (orange) and $5$ femtoseconds (green). The dashed lines represent the Fermi-liquid approximation~\eqref{eq:FLT}. (b) The {\em e}-{\em e} relaxation rate at $\e = \e_F$ (blue dots) and $\e = \e_F + 1 eV$ (orange dots) as a function of $|E_0|^2$. (c) The corresponding results for the {\em e}-{\em ph} relaxation rate; all three lines are indistinguishable.
}\label{fig:eta}
\end{figure}

Having understood the early excitation stage, we turn to the consequent relaxation stage. Figure~\ref{fig:eta}(a) shows the dynamics of the electron energy dependent {\em e}-{\em e} collision (i.e., thermalization) rate, $\eta_\textrm{e-e}(\e,t)$, given by the functional derivative of the corresponding {\em e}-{\em e} interaction term~\cite{Quantum-Liquid-Coleman,Un-Sarkar-Sivan-LEDD-I,Un-Sarkar-Sivan-LEDD-II,single_cycle_nlty_Article}. In the language of atomic and molecular physics, this behaviour corresponds to a time- and energy-dependent depopulation time ($T_1$) and dephasing time ($T_2$). Remarkably, this dependence can be reproduced by an analytic form, which is a generalization of FLT. In particular, in Section~\ref{app:generalized_FLT}, we show that 
\begin{align}\label{eq:FLT}
\eta^T_\textrm{e-e}(\e,\tilde{T}_e) \approx K(\tilde{T}_e) \Bigg[A_0(\e,\tilde{T}_e) (k_B\tilde{T}_e)^2 
+ A_1(\e,\tilde{T}_e) (k_B\tilde{T}_e)\left(\e-\mu(\tilde{T}_e)\right)
+ A_2(\e,\tilde{T}_e)\left(\e-\mu(\tilde{T}_e)\right)^2\Bigg],
\end{align} 
where $k_B$ is the Boltzmann constant, $\tilde{T}_e = \tilde{T}_e(t)$ is, again, a time-dependent effective electron temperature (extracted from the first energy moment of the electron distribution, i.e., the diagonal elements of the DM~\cite{delFatti_nonequilib_2000,Italians_hot_es,GdA_hot_es,Un-Sarkar-Sivan-LEDD-I}), $\mu(\tilde{T}_e)$ is the chemical potential~\cite[Equation~(24)]{Un-Sarkar-Sivan-LEDD-I}, the prefactor $K(\tilde{T}_e)$ is a generalization of the standard FLT expression for noble metals~\cite{Quantum-Liquid-Coleman} to the one given in Equation~(S25)
or Equation~(19) of Reference~\cite{Un-Sarkar-Sivan-LEDD-I} for TCOs, and the coefficients $A_{0,1,2}(\e,\tilde{T}_e)$ (see Equation~(S26)
) include the correction for high temperatures resulting from the low electron density, hence, the smallness of $\mu(\tilde{T}_e)$. 

Specifically, one can observe that the {\em e}-{\em e} collision rate grows with the amount of energy injected into the electron system via photon absorption. As in standard FLT~\cite{Quantum-Liquid-Coleman}, electrons far from the Fermi energy undergo collisions more frequently than electrons close to the Fermi energy, yet the new linear correction we derived (see Section~S2
) {\em partially suppresses} this effect. We also find, in similarity to standard FLT, that the dip of the {\em e}-{\em e} collision rate for electrons near the Fermi energy shifts from $\e_F$ to $\mu(\tilde{T}_e)$ and is gradually washed out with increasing (effective) electron temperature (which reaches $\sim 4500$ K for the incident illumination intensity of $1$ TW/cm$^2$ used in this example), such that the thermalization time becomes nearly energy-independent, with a value of $\sim 12$ fs. The success of the generalized FLT expression~\eqref{eq:FLT} to reproduce the rigorous thermalization dynamics also means that the deviation from thermal equilibrium per se has a small effect on the {\em e}-{\em e} collision rate, see Figure~\ref{fig:eta}(a). This is remarkable, since the derivation of FLT is performed under the condition that the electron system is nearly thermalized, a condition which clearly does not hold here (Figure~\ref{fig:fe_0.6_1.2_2.4Vm-1}).

Figure~\ref{fig:eta}(b) further shows that the analytic approximation for $\eta_\textrm{e-e}$ scales linearly with the illumination intensity, with electrons of different energies exhibiting somewhat different slopes. Assuming this scaling persists for higher intensities, it is likely that {\em e}-{\em e} collision rates on the scale of a single femtosecond would be reached for intensities of $10$ TW/cm$^2$, which are in use for many high-harmonic generation experiments. In that sense, our results a-posteriori support the choices of fixed $T_2$ values of single femtoseconds~\cite{Huber_T2_1fs,Goulielmakis_T2_1fs,Luu_PRB_2016_T2_1fs,Vampa_PRL_2014,Lu_PRA_2016_T2_1fs,Ghimire_T2_1fs,Yabana_PRA_T2_diamond} shown to yield a good match with high-harmonic generation experiments, at least after the initial excitation stage.

In that respect, it is also worth mentioning that the relaxation behaviour we reveal may be different when exciting from the valence band~\cite{Stockman-current-in-dielectrics,Stockman-current-in-dielectrics2} or due to tunnelling under a voltage bias~\cite{Aizpurua-Greffet,Aizpurua_Borissov_Kruger_tunnelling_ACS_Photonics_2025}. A comprehensive study of the relaxation dynamics in all these systems will be the topic of future investigations.

While {\em e}-{\em e} collisions cause thermalization, only the Umklapp fraction contributes to the permittivity rise~\cite{single_cycle_nlty_Article}, whereas the permittivity relaxation is determined by the {\em e}-{\em ph} collision time, $\eta_\textrm{e-ph}$. This is a natural consequence of the dependence of the permittivity change $\Delta \varepsilon$ on the increase of the electron energy (i.e., in proportion to the (net) absorption). Figure~\ref{fig:eta}(c) shows that the {\em e}-{\em ph} collision rate is essentially independent of the dynamics/electron energy/distribution. This happens because the phonon energy is relatively small (below the Debye energy of 0.06 eV~\cite{ITO_properties_2016}), making $\eta_\textrm{e-ph}$ primarily governed by the average phonon number (see also Equation~(13) of Reference~\cite{Un-Sarkar-Sivan-LEDD-II}). Furthermore, Figure~\ref{fig:eta}(c) shows that $\eta_\textrm{e-ph}$ is much slower than $\eta_\textrm{e-e}$, hence, {\em e}-{\em ph} interactions do not explain the rapid decay observed in Reference~\cite{Shalaev_Segev_nanophotonics_2023}. \XYZ{This also implies that the nonlinearity will be accumulated between consequent pulse incidence, so that application of an intense periodic illumination for the purpose of generating a photonic time crystal is likely to quickly bring the ITO to the damage threshold after a few oscillations, making the realization of photonic time crystal in such materials unlikely.}{This slow relaxation also has important implications for the realization of optical-frequency photonic time crystals (PTCs) in TCOs~\cite{Segev_Shalaev_PTCs,Khurgin_photonic_time_crystal_vs_4Wmix,Rockstuhl_PTCs}. Since optical/NIR PTCs require modulation periods on the order of a few femtoseconds, energy deposition occurs on timescales much shorter than those associated with thermal diffusion or heat extraction to the substrate\footnote{\XYZ{}{For example, using a thermal diffusivity of $\sim100$ cm$^2$/s~\cite{ICFO_Sivan_metal_diffusion} and a characteristic length scale of $\sim100$ nm, the thermal diffusion time is on the order of hundreds of picoseconds.}}. 
The cumulative thermal nonlinearity identified in this work consequently imposes a strong constraint on TCO-based optical PTCs, as repeated modulation cycles would progressively increase the energy of the electron system and could rapidly drive the material toward its damage threshold. This argument does not preclude PTC operation at lower frequencies, where the modulation timescale is sufficiently long to allow heat dissipation, as demonstrated experimentally in the THz regime~\cite{THz_TPCs_2026}.}

This behaviour contrasts the instantaneous SiO$_2$ nonlinearity studied in Reference~\cite{Schultze_Kerr_SiO2} which involved mostly virtual transitions from the valence to the conduction band~\cite{Schultze_Kerr_SiO2,Narimanov_2025} (so that relaxation is inherent) rather than real transitions within the conduction band (for which the relaxation occurs via {\em e}-{\em ph} interactions); this reflects, again, the absorptive (or, in the language of Reference~\cite{Schultze_Kerr_SiO2}, irreversible) nature of the intraband nonlinearity of TCOs studied here.

\section{Conclusion and outlook}
Our calculations yield a nonlinear response which is at least an order of magnitude stronger than the prediction in Reference~\cite{Narimanov_2025} for the Kerr (interband) nonlinearity and much larger than the experimentally measured transmission and reflection changes in Reference~\cite{Shalaev_Segev_nanophotonics_2023}. Our prediction for the decay dynamics is also at odds with the observed 40-femtosecond decay time of the transmission. Our theory could nevertheless be reconciled with that measurement if, in the experiment, the electron system is brought closer to transparency than in our simulations, an effect which may then expose the weaker and instantaneous Kerr nonlinearity considered in Reference~\cite{Narimanov_2025} and measured in Reference~\cite{Schultze_Kerr_SiO2}\XYZ{}{ in silica}. This could happen for a different set of material parameters and/or much higher intensities than used here; It may also originate from multi-photon absorption (which is more likely in the configuration of Reference~\cite{Shalaev_Segev_nanophotonics_2023} than in our configuration), which was shown to cause a partial cancellation of the intraband nonlinearity we studied~\cite{Clerici_Nat_Comm_2017}. Yet, such mechanisms do not explain the far slower decay of the reflection observed in Reference~\cite{Shalaev_Segev_nanophotonics_2023}. It is also still unknown how the Kerr nonlinearity would be affected by the drastic population changes and extremely high electron temperatures; a proper evaluation of these effects requires a comprehensive model that accounts for both interband and intraband transitions. 
Hints about the contributions of these effects, and many additional nonlinear mechanisms, can be extracted from the comprehensive hydrodynamic model used in Reference~\cite{Scalora-ITO-2025} to reproduce the experimental data of Reference~\cite{Shalaev_Segev_nanophotonics_2023}. 
Extending that approach to include the temperature and effective-mass dynamics unique to ITO may indeed yield a better match to the experimental data. 

\XYZ{}{In addition, the 4 nm sphere considered in the present work supports nearly uniform electric fields and electron temperatures. This differs from the 300 nm ITO film studied in Reference~\cite{Shalaev_Segev_nanophotonics_2023}, where substantial field and temperature gradients may develop, contributing to the measured response through transport and spatiotemporal effects (some of which are discussed in~\cite{Scalora-ITO-2025} as well). These additional effects may therefore play a role in the experimental response and could contribute to the discrepancy between our predictions and the measured dynamics.}

\XYZ{}{Furthermore, the effect of nonlocality~\cite{Stephane-ITO-nonlocality-2025} which arises in few nanometer structures such as those we study may give rise to a size-dependence the plasma frequency and modify the resonance position and local-field enhancement~\cite{DTU_nonlocal,Baumberg-Aizpurua,Scalora-ITO-2020,Bondarev_PRR_2020,Ciraci_PRX_2021}. These effects are orthogonal to the main findings in the paper, and are expected to affect the results in this work, at most, in a modest quantitative manner.}

\XYZ{}{Finally, fieldoscopy measurements have resolved the transient electric-field in ITO nanocrystals driven by two-cycle optical pulses~\cite{Herbst_fieldoscopy_AdvSci_2026}. That work focused on the reversibility of the macroscopic ultrafast response, its intensity scaling and recovery time, thus, providing the previously missing experimental route for directly testing the microscopic electron dynamics and nonlinear response predicted in this work.}

\medskip
\noindent \textbf{Supporting Information} \par 
\noindent Supporting Information is available from the Wiley Online Library or from the author.

\medskip
\noindent \textbf{Acknowledgements} \par 
\noindent I.W.U. was funded by the Guangdong Natural Science Foundation (Grants No. 2024A1515011457). Y.S. was partially funded by a Lower-Saxony-Israel collaboration grant no. 76251-99-7/20 (ZN 3637) as well as an Israel Science Foundation (ISF) grant (340/2020).

\medskip

%

\end{document}